\documentclass[journal]{IEEEtran}
\usepackage{latexsym}
\usepackage{graphicx}
\usepackage{amsfonts,amssymb,amsmath}
\usepackage[utf8]{inputenc}
\usepackage{graphicx}
\usepackage{cite}
\usepackage{amsmath,amsthm,amssymb,mathrsfs}
\usepackage{epstopdf}
\usepackage{threeparttable}
\usepackage{multirow}
\usepackage[colorlinks, citecolor=blue]{hyperref}
\usepackage{caption}
\usepackage{mathtools}
\usepackage{lscape}

\usepackage{float}
\usepackage{subfig}
\usepackage[table]{xcolor}
\usepackage{stfloats}
\usepackage{booktabs} 

\usepackage{colortbl}
\definecolor{tabcolor}{rgb}{.105,.410,.113}

\def\BibTeX{{\rm B\kern-.05em{\sc i\kern-.025em b}\kern-.08em T\kern-.1667em\lower.7ex\hbox{E}\kern-.125emX}}
\begin{document}

\title{An Overview of Generic Tools for Information-Theoretic Secrecy Performance Analysis over Wiretap Fading Channels}

\author{Long~Kong, Yun~Ai,~\IEEEmembership{Member,~IEEE}, Lei~Lei,~\IEEEmembership{Member,~IEEE},~Georges~Kaddoum,~\IEEEmembership{Member,~IEEE},\\~Symeon Chatzinotas,~\IEEEmembership{Senior Member,~IEEE}, and Björn Ottersten,~\IEEEmembership{Fellow,~IEEE}

\thanks{This work was supported in part by the Luxembourg National Research Fund (FNR) projects, titled Exploiting Interference for Physical Layer Security in 5G Networks (CIPHY), and in part by the Energy and CompLexity EffiCienT mIllimeter-wave Large-Array Communications (ECLECTIC).}
\thanks{L. Kong, L. Lei, S. Chatzinotas, and B. Ottersten are with the Interdisciplinary Centre for Security, Reliability and Trust (SnT), University of Luxembourg, Luxembourg City 1855, Luxembourg (e-mail: \{long.kong, lei.lei, symeon.chatzinotas, björn.ottersten\}@uni.lu.).}
\thanks{Y. Ai is with the Faculty of Engineering, Norwegian University of Science and Technology, 2815 Gjøvik, Norway (e-mail: yun.ai@ntnu.no).}
\thanks{G. Kaddoum is with the LaCIME Laboratory, Department of Electrical Engineering, École de Technologie Supérieure (ÉTS), Université du Québec, Montréal (Québec), Canada, H3C 1K3, e-mail: (georges.kaddoum@etsmtl.ca.).}}
\maketitle

\begin{abstract}
An alternative or supplementary approach named as physical layer security has been proposed to afford an extra security layer on top of the conventional cryptography technique. In this paper, an overview of secrecy performance investigations over the classic Alice-Bob-Eve wiretap fading channels is conducted. On the basis of the classic wiretap channel model, we have comprehensively listed and thereafter compared the existing works on physical layer secrecy analysis considering the small-scale, large-scale, composite, and cascaded fading channel models. Exact secrecy metrics expressions, including secrecy outage probability (SOP), the probability of non-zero secrecy capacity (PNZ), average secrecy capacity (ASC), and secrecy bounds, including the lower bound of SOP and ergodic secrecy capacity, are presented. In order to encompass the aforementioned four kinds of fading channel models with a more \textit{generic} and \textit{flexible} distribution, the mixture gamma (MG), mixture of Gaussian (MoG), and Fox's $H$-function distributions are three useful candidates to largely include the above-mentioned four kinds of fading channel models. It is shown that they are flexible and general when assisting the secrecy analysis to obtain closed-form expressions. Their advantages and limitations are also highlighted. Conclusively, these three approaches are proven to provide a unified secrecy analysis framework and can cover all types of independent wiretap fading channel models. Apart from those, revisiting the existing secrecy enhancement techniques based on our system configuration, the on-off transmission scheme, jamming approach (including artificial noise (AN) \& artificial fast fading (AFF)), antenna selection, and security region are presented.  
\end{abstract}


\maketitle

\section{INTRODUCTION}
\IEEEPARstart{T}{he} exposure of confidential messages into the wireless transmission medium makes the legitimate transmission vulnerable due to (i) the openness of wireless transmission medium; and (ii) the standardization of wireless transmission schemes, e.g., coding and modulation schemes. The conventionally widely used approach called the cryptography technique is placed at the upper layer under the critical assumptions of (i) error-free links at physical layer; and (ii) incapability of eavesdroppers to decrypt the secret key due to limited computational power. It is predicted that in the near future the quantum computing will be able to easily break the current strong public-key cryptosystems \cite{8967098}. Besides, issues like computational complexity and key distribution and management in some emerging decentralized networks (e.g., sensor or radio-frequency identification (RFID) networks) makes it difficult to deploy public-key infrastructure \cite{Poor19}. Against this background, numerous works suggest that a security additional be achieved by making full use of the impairments (i.e., noise and fading) of wireless communication links \cite{zhou2013physical}. In other words, under the cover of noise and interference, private messages can be securely transmitted to the legitimate receivers at the physical layer in the presence of an unauthorized devices or a potential malicious eavesdropper. Consequently, research efforts were shifted to seek a cost-efficient and effective solution to address the secrecy concern from the physical layer of the layer-structure protocol. 

An appealing countermeasure, known as physical layer security (PLS), was found suitable for preventing eavesdropping attacks on wireless communication from the information-theoretic perspective. Two pioneering works were respectively laid by Shannon \cite{Shan49} and Wyner \cite{6772207}, where the notion of perfect secrecy and wiretap channel model were respectively introduced. It is noteworthy to point out that Wyner's result established the PLS from the system model level, where he considered the three-user scenario, consisting of a source (Alice), an intended legitimate user (Bob), and an eavesdropper (Eve) over the discrete memoryless wiretap channel. In \cite{Leung78}, Wyner's wiretap model was extended to the Gaussian wiretap channel, Leung \textit{et al.} also defined the secrecy capacity as the difference between the channel capacity of the main channel (Alice to Bob) and that of the wiretap channel (Alice to Eve). The conceptual beauty of secrecy capacity indicates that only when the legitimate link experiences better quality of received signals compared to the wiretap channel, positive secrecy can thereafter be surely guaranteed. In the last few decades, a growing body of secrecy metrics investigations over wiretap fading channels were conducted, e.g., \cite{4529264,4626059}. The insights drawn from these works offer mathematical proofs that the fading property of wireless channels can be reversely used to enhance secrecy. To this end, various researchers from both the wireless communication and signal processing communities were inspired to explore this effective secrecy enhancement solution.

Observing the existing surveys and tutorial works related to the PLS~\cite{Poor19,5751298,6739367,7467419,7762075,8335290,8850067}, techniques like information-theoretic security, artificial-noise aided security, security-oriented beamforming, security diversity methods, and physical layer secret key generation are listed in \cite{7467419} by Zou \textit{et al}. As an organic part of PLS techniques, information-theoretic security have been classified into three categories: (i) memoryless wiretap channels; (ii) Gaussian wiretap channels; and (iii) fading wiretap channels, however, the majority of information-theoretic security is centered around the fading wiretap channels, see references \cite{4529264,4626059,6698305}. To be specific, Bloch \textit{et al.} in \cite{4529264} examined the impact of the fading property of wireless channels on the secrecy issue and thereafter proposed the feasibility of the average secure communication rate and the outage probability as secrecy metrics. Later on, Gopala \textit{et al.} in \cite{4626059} established the perfect secrecy capacity over the fading wiretap channel model when (i) the full channel state information (CSI) are available at the transmitter; and (ii) only the main channel CSI is perfectly known at the transmitter. In addition, the on/off power allocation strategy was proposed as a transmission policy, i.e., Alice can perform information transmission as long as the channel gain of the legitimate user is larger than a predetermined positive threshold. 

Since then, numerous works have analyzed the secrecy performance over a diverse body of fading channels, just to name some, Rayleigh \cite{4529264}, Nakagami-$m$, Weibull \cite{6831147}, Rician (Nakagami-$q$) \cite{6338984}, Hoyt (Nakagami-$n$) \cite{8281715,8672771}, Lognormal \cite{7058442}, $\alpha-\mu$ (equivalently generalized gamma) \cite{7094262,7374839,7856980,OutageLongLett,8477185}, $\kappa-\mu$ \cite{6957529,7467556,7996664,8013132}, $\eta-\mu$ \cite{8500351}, generalized-$\mathcal{K}$ \cite{7313027,7342908,7654915,8279403}, extend generalized-$\mathcal{K}$ (EGK) \cite{8706707}, Fisher-Snedecor $\mathcal{F}$ \cite{LongFisherF}, gamma-gamma \cite{8358703}, shadowed $\kappa-\mu$ \cite{snchez2020physical}, double shadowed Rician \cite{8745490}, Fox's $H$-function \cite{8706707}, cascaded Rayleigh/Nakagami-$m$/$\alpha-\mu$ \cite{8403278,8441332,8354927}, $\alpha-\kappa-\mu$/$\alpha-\eta-\mu$ \cite{8556482}, Beaulieu-Xie \cite{CHAUHAN2020152940}, $\alpha-\kappa-\eta-\mu$ \cite{8543055,8421203}, two-wave with diffuse power (TWDP) \cite{6698305}, $N$-wave with diffuse power (NWDP) \cite{snchez2020secrecy}, $\kappa-\mu$/Gamma \cite{8929467}, Fluctuating Beckmann \cite{8813034}, correlated Rayleigh \cite{5730598}, correlated $\alpha-\mu$ \cite{8727439}, correlated shadowed $\kappa-\mu$ \cite{8746270}, mixed $\eta-\mu$ and Málaga \cite{8501589}, Málaga \cite{7859265,8422835}, fluctuating two-ray (FTR) channels \cite{8369134,8777365}. The usage of these fading channels are examined practical and feasible in various wireless communications, such as, cellular device-to-device (D2D), vehicle-to-vehicle (V2V) communications \cite{7467556}, mmWave communications \cite{8369134}, underwater acoustic communications (UAC), frequency diverse array (FDA) communications \cite{8672807}, body-centric fading channels, unmanned aerial vehicle (UAV) systems, land mobile satellite (LMS), etc. \cite{8002635,snchez2020physical}.

To the authors' best knowledge, no survey papers so far have focused on the secrecy metrics analysis over various fading channels. To this end, the main contributions of this work are listed as follows: 
\begin{itemize}
\item[1)] presenting the state-of-the-art of information-theoretic secrecy analysis works over four kinds of wiretap fading models, (i) small-scale, (ii) large-scale, (iii) composite, and (iv) cascaded.
\item[2)] summarizing and comparing three useful secrecy analysis approaches, i.e., mixture gamma (MG), mixture of Gaussian (MoG), and Fox's $H$-function based solutions. The insights drawn herein show that these three tools are useful to encompass most existing secrecy analysis works by properly configuring the fading channel characteristics in the manner of the mathematical expressions of the three tools.
\item[3)] providing a list set of secrecy enhancement solutions, including the on-off transmission scheme, artificial noise (AN) and artificial fast fading (AFF), jamming approach, antenna selection, and security region for the classical wiretap channel model.
\end{itemize} 

The remainder of this paper is organized as follows: Section \ref{Sec2} presents the classic wiretap model and the problem formulation. In \ref{Sec3}, we divide the secrecy metrics analysis works into four categories according to the fading channel model and consequently present three useful and proven tools used to assist the secrecy metrics analysis. In Section \ref{Sec4}, we introduce the secrecy enhancement schemes based on classical wiretap fading channels. Finally, Section \ref{Sec5} concludes this paper.
\section{System model and problem formulation} \label{Sec2}
\subsection{System Model}
Consider the classic Alice-Bob-Eve wiretap channel model, shown in Fig. 1, where a transmitter (Alice) intends to send confidential messages to the legitimate receiver (Bob) in the presence of a malicious eavesdropper (Eve). The link between Alice and Bob is called the main channel while the one between Alice and Eve is the wiretap channel. The instantaneous signal-to-noise ratio (SNR) at Bob ($B$) and Eve ($E$) is expressed as $\gamma_i = \bar{\gamma}_i g_i, i \in \{B,E\}$, where $\bar{\gamma}_i$ is the average received SNR and $g_i$ is the channel gain, which can be modeled by any of the aforementioned fading distributions.
\begin{figure}[!t]
\centering{\includegraphics[width=0.8\columnwidth]{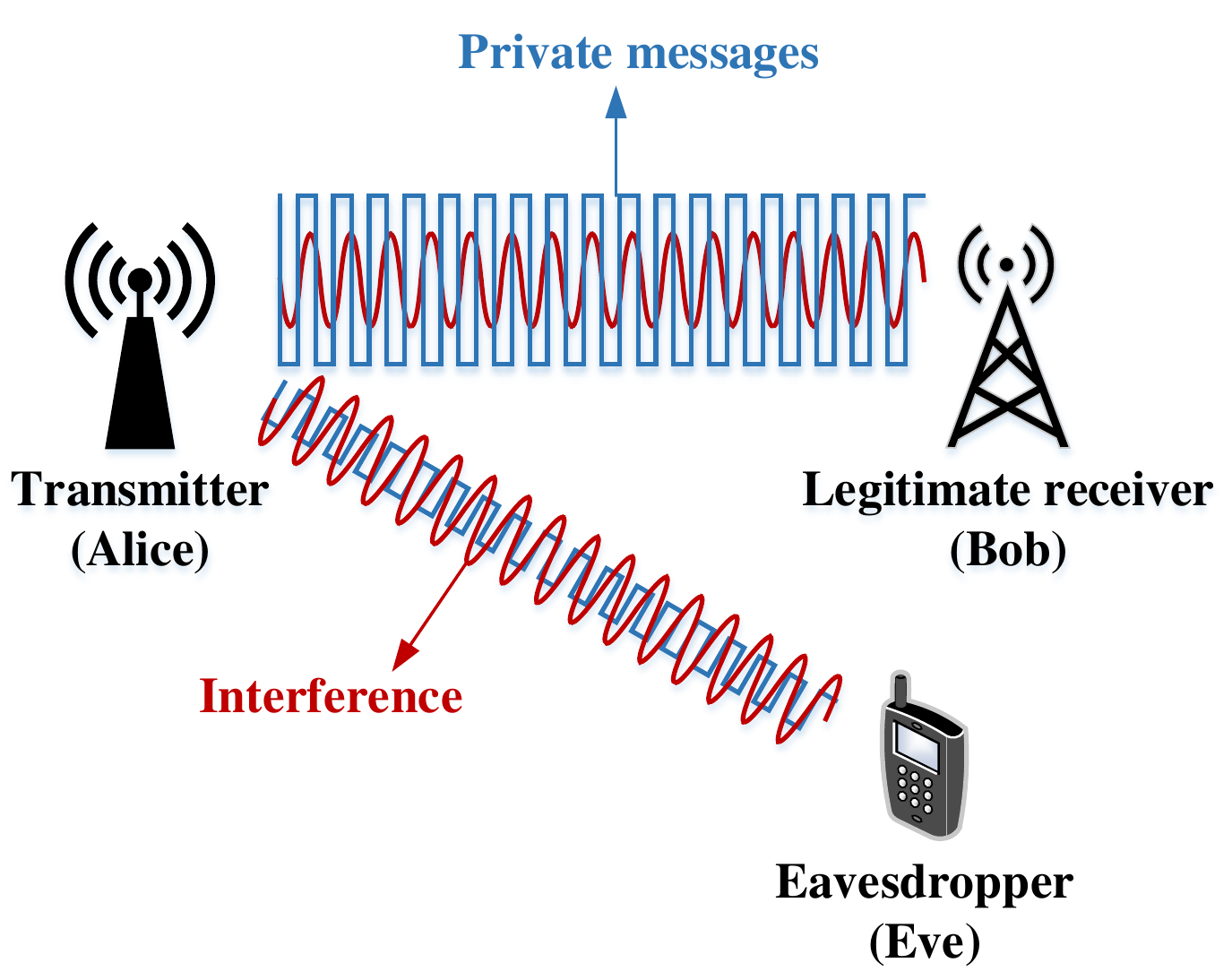}}
\caption{A three-node wireless wiretap system model.}
\label{SysMedol}
\end{figure}
\subsection{Secrecy metrics}
Assuming the availability of perfect CSI at all terminals. According to \cite{4529264}, the instantaneous secrecy capacity for one realization of the ($\gamma_B, \gamma_E$) pair over quasi-static wiretap fading channels is given by 
\begin{equation}
C_s(\gamma_B,\gamma_E) = \left[\underbrace{\log_2\left(1+\gamma_B\right)}_{C_M}-\underbrace{\log_2\left(1+\gamma_E\right)}_{C_W}\right]^+,
\end{equation}
where $[x]^+ \overset{\triangle}{=} \max(x,0)$. $C_M$ and $C_W$ are the channel capacity of main and wiretap channels, respectively.

Based on the definition of the instantaneous secrecy capacity, secrecy metrics including the secrecy outage probability (SOP), the probability of non-zero secrecy capacity (PNZ), average secrecy capacity (ASC), and ergodic secrecy capacity are henceforth developed for the sake of evaluating the security of all kinds of wireless systems.
\subsubsection{Secrecy outage probability (SOP)} 
Considering the passive eavesdropping scenario, where Alice transmits private messages at a constant secrecy rate $R_t$ whilst in the presence of a passive Eve, who only listens to the main channel without sending any probing messages. Against this background, perfect secrecy can be assured only when $R_t$ falls below the instantaneous secrecy capacity $C_s$. Strikingly, the SOP metric is commonly seen as a crucial secrecy indicator to measure the level that how perfect secrecy is compromised. Mathematically speaking, the SOP is the probability that the instantaneous secrecy capacity is lower than a predetermined secrecy rate $R_t$, 
\begin{equation}
\mathcal{P}_{out}  = Pr\left( C_s \le R_t\right).
\end{equation}
\subsubsection{The probability of non-zero secrecy capacity (PNZ)}
The PNZ is regarded as another important secrecy metric that measures the existence of positive secrecy capacity with a probability, 
\begin{equation} \label{PNZ_def}
\begin{split}
\mathcal{P}_{nz}& = Pr \left( C_s  > 0\right) \\
& \mathop=^{R_t=0} 1- Pr(Cs \le R_t)\\
&= Pr(\gamma_B > \gamma_E).
\end{split}
\end{equation}
From (\ref{PNZ_def}), usually, the PNZ metric can be easily obtained from the SOP metric by setting $R_t =0$.
\subsubsection{Average secrecy capacity (ASC)}
When an active eavesdropper appears, the average secrecy capacity serves as a useful measurement that guides Alice to adapt its transmission rate based on $C_M$ and $C_N$ so as to achieve perfect secrecy. In other words, the ASC is a secrecy metric that evaluates how much achievable secrecy rate can be guaranteed for the whole system. It is mathematically defined as
\begin{equation}
\bar{\mathcal{C}} = \mathcal{E}[C_s(\gamma_B,\gamma_E)],
\end{equation}
 where $\mathcal{E}[\cdot]$ is the expectation operator.
\subsubsection{Ergodic secrecy capacity}
As an appropriate secrecy measure to characterize the time-varying feature of wireless channels, the ergodic secrecy capacity is resultantly utilized to quantify the ergodic features of wireless channels \cite{5699835,6623091,LongIWCMC2018,7247765,8477185}. The ergodic secrecy capacity is mathematically evaluated by averaging the channel capacity over all fading channel realizations, which is computed as 
\begin{equation}
\mathcal{E}(C_s) = \left[\mathcal{E}[\log_2(1+\gamma_B)] - \mathcal{E}[\log_2(1+\gamma_E)\right]^+.
\end{equation}

\section{Secrecy Characterization} \label{Sec3}
\subsection{Exact secrecy analysis}
In wireless communication systems, the transmitted signals are reflected, diffracted, and scattered from objects that are present on their path to the receiver. The received signals experience fading (multipath) and shadowing (signal power attenuation or pathloss) phenomena, which pose destructive and harmful impacts at the receiver sides. The essence of PLS is to reversely use the impairments of wireless channels as secrecy enhancement means. Under the assumption that the main and wiretap channels undergo independent fading conditions. This section mainly presents the secrecy analysis over wiretap fading channels according to the existing fading channel categories.
\subsubsection{Small-scale fading channels}
The random changes in signal amplitude and phase from the spatial positioning between a receiver and a transmitter is referred to small-scale fading. The well-known small-scale fading models are Rayleigh, Nakagami-$m$, Rician, $\alpha-\mu$ (equivalently, generalized gamma or Stacy), etc. The simple and tractable form of these models makes small-scale fading appealing and popular in the security and reliability performance analysis. Examples can be found in \cite{4529264,7094262,7374839,7856980,6338984,OutageLongLett}, where SOP, PNZ, and ASC are analyzed by devising either closed-form or highly tight approximated expressions. It is noteworthy of mentioning that the $\alpha-\mu$ distribution can be reduced to Rayleigh ($\alpha=2,\mu=1$), Nakagami-$m$ ($\alpha =2,\mu=m$), Weibull ($\alpha$ is the fading parameter, $\mu =1$), and gamma ($\alpha=1$, $\mu$ is the fading parameter) distributions by properly attributing the values of $\alpha$ and $\mu$. To this end, the applicability and flexibility of the $\alpha-\mu$ distribution have been well explored in the literature. In addition, the TWDP fading model is also of high flexibility as it includes Rayleigh, Rician, and hyper-Rayleigh fading as special cases. Besides, it can also model a link worse than Rayleigh fading. It characterizes propagating scenarios where the received signal contains two strong, specular multipath waves. The PLS investigation over TWDP wiretap fading channels was studied in \cite{6698305}. Apart from the aforementioned works, in \cite{7543509}, the authors studied the effect of eavesdroppers' location uncertainty on the SOP metric, where Eve is supposed to be located in a ring-shaped area around Alice and undergoes Rayleigh fading.

Another interesting direction of PLS over small-scale fading channels lies in the secrecy investigation over correlated fading channels. The correlation is caused due to the distances between Bob and Eve, or the scattering environments, etc. The physical correlation essentially makes the fading statistics, i.e., the mathematical representation of the joint PDF of $\gamma_B$ and $\gamma_E$, fairly complex and eventually makes it intractable and highly difficult to obtain exact closed-form secrecy performances, instead, secrecy bounds are derived (see references \cite{5730598,8727439}).
\subsubsection{Large-scale fading channels} 
The so-called large-scale fading results from signal attenuation due to signal propagation over large distances and diffraction around large objects (i.e., hills, mountains, forests, billboards, buildings, etc.) in the propagation path. One widely studied example of large-scale fading channels is the lornormal distribution, however its complex mathematical form hinders the derivation of exact reliability and secrecy performance expressions. For instance, Pan \textit{et al.} in \cite{7058442} investigated the PLS over non-small scale fading channels, wherein independent/correlated lognormal fading channels and composite fading channels were considered and highly accurate approximated secrecy representations were derived. 
\subsubsection{Composite fading channels}
Different from the small-scale (fading) and large-scale (shadowing) fading models, composite fading models are proposed to account for the effects of both small-scale and large-scale fading simultaneously. For instance, Kumar \textit{et al.} in \cite{7467556} presented the SOP, PNZ, and ASC over $\kappa-\mu$ fading channels and explored the obtained results in a diverse range of wireless communication scenarios, including cellular D2D, body area networks (BAN), and V2V. Moualeu and Hamouda in \cite{8013132} subsequently extended the results in \cite{7467556} to the single-input multiple-output (SIMO) scenario and derived the ASC and lower bound of SOP. More recently, to elaborate the shadowing effect of wireless channels, in \cite{8745490,8746270}, the authors investigated the secrecy performance over the shadowed Rician and $\kappa-\mu$ wiretap fading channels.

Other widely used models, e.g., Rayleigh/Lognormal (RL), Nakagami-$m$/Lognormal (NL), generalized-$\mathcal{K}$, gamma-gamma, and Fisher-Snedecor $\mathcal{F}$, are examined in realistic wireless communication scenarios to model the channel-induced physical layer dynamics. To be specific, the Fisher-Snedecor $\mathcal{F}$ fading model was proposed to characterize device-to-device (D2D) communications, where its simplicity and feasibility are compared with the generalized-$\mathcal{K}$ fading model in \cite{7886273}. Similarly, the gamma-gamma, mixed $\eta$–$\mu$ and Málaga, Málaga distributions were shown feasible to accurately model the radio frequency-free space optical (RF-FSO) links, and the secrecy performance over those fading models are respectively provided in \cite{7968415,8501589,7859265,8422835}. To encompass more special fading models in one distribution, secrecy analysis over $\alpha-\eta-\mu$, $\alpha-\kappa-\mu$, and $\alpha-\eta-\kappa-\mu$ fading models is investigated in \cite{8556482,8421203,8543055}. For example, the $\alpha-\eta-\kappa-\mu$ model can be realized to the Rayleigh, Nakagami-$m$, Rician, $\kappa-\mu$, $\eta-\mu$, $\alpha-\mu$, etc. Those models are highly general and flexible, however its complex mathematical representation of characteristics makes it difficult to derive the exact closed-form secrecy metrics.
\subsubsection{Cascaded fading channels}
Cascaded fading models were found feasible to characterize the multi-hop non-regenerative amplify-and-forward (AF) relaying with fixed gain, the propagation in the presence of keyholes, and the keyhole/pinhole phenomena in MIMO systems, as well as the quite recent proposed intelligent reflective surface (IRS) scenario \cite{9134962}. Yang \textit{et al}. in \cite{9134962} modeled the IRS-aided main link as a multiplication of two Rayleigh distributed random variable. For vehicular networks, the secrecy performance has been investigated considering the double Rayleigh fading channels. For other works over cascaded Nakagami-$m$/Fisher-Snedecor $\mathcal{F}$/$\alpha-\mu$ wiretap fading channels, the readers are referred to \cite{8403278,8354927,8441332,8629417,8877263}. As discussed earlier, the cascaded $\alpha-\mu$ fading channel similarly includes the cascaded Rayleigh, cascaded Nakagami-$m$, cascaded Weibull, and cascaded gamma distributions. We have in \cite{8354927} studied the SOP, PNZ, and ASC performances with closed-form expressions, which are given in terms of Fox's $H$-function. The obtained results therein are identical to the exact analytical representations given in \cite{8629417,8877263}.

The majority of information-theoretic security analysis works over wiretap fading channels are summarized and their contributions are highlighted in Table \ref{table1}.
\begin{table}[!t]
\renewcommand{\arraystretch}{1.3}
\caption{\textsc{Major information-theoretic secrecy analysis works over the classical wiretap fading channels }}
\label{table1}
\centering
\begin{tabular}{|p{0.6cm}|p{2.1cm}|p{4.45cm}|}
\arrayrulecolor{tabcolor} \toprule[1.4pt] 
\hline
\rowcolor{gray}
\textbf{Year} &\textbf{References}& \textbf{Contributions}\\ \hline
2008&Bloch \textit{et al.} \cite{4529264}& derived simple and exact SOP, PNZ, and ASC closed-form expressions over \textbf{Rayleigh} fading channels \\ \hline
2013& Liu \cite{6338984,6831147}& derived the PNZ over \textbf{Rician} and \textbf{Weibull} fading channels\\ \hline
2014 & Wang \textit{et al.} \cite{6698305}& derived the ASC and SOP over \textbf{TWDP} fading channels \\ \hline
2015-2018& Lei \textit{et al.} \cite{7094262,7856980}, Kong \textit{et al.} \cite{7374839,OutageLongLett}& analyzed the SOP, lower bound of SOP, PNZ, and ASC over $\alpha-\mu$ wiretap fading channels. \\ \hline
\multirow{3}{*}{2016} & Pan \textit{et al.} \cite{7058442}& proposed an highly accurate approximated secrecy solution over the \textbf{lognormal} fading channels.\\ \cline{2-3}
 &Bhargav \textit{et al.}\cite{7467556}& derived the lower bound of SOP and PNZ over $\kappa-\mu$ fading channels.\\ \cline{2-3}
 & Lei \textit{et al.}\cite{7313027,7342908,7654915}& analyzed the secrecy metrics over \textbf{generalized}-$\mathcal{K}$ fading channels.\\ \hline
 2017 & Saber and Sadough \cite{7859265} & derived the SOP, PNZ, and ASC over the \textbf{Málaga} wiretap fading channels.\\ \hline
\multirow{3}{*}{2018} & Kong \& Kaddoum\cite{LongFisherF}  & derived the SOP, lower bound of SOP, PNZ and ASC over \textbf{Fisher-Snedecor} $\mathcal{F}$ wiretap fading channels.\\ \cline{2-3}
& Kong \textit{et al.} \cite{8354927}& derived closed-form expressions for the SOP, PNZ, and ASC over \textbf{cascaded} $\alpha-\mu$ wiretap fading channels. \\ \cline{2-3}
& Mathur \textit{et al.} \cite{8421203} & derived the ASC and SOP over $\alpha-\eta-\kappa-\mu$ wiretap fading channels. \\ \hline
\multirow{3}{*}{2019} & Kong \& Kaddoum \cite{8672771}& analyzed the secrecy metrics with the assistance of the \textbf{MG} distribution. \\ \cline{2-3}
 & Kong \textit{et al.}\cite{8706707}& analyzed the secrecy metrics over a general and flexible \textbf{Fox's $H$-function} wiretap fading channels.\\ \cline{2-3}
 &Moualeu \textit{et al.} \cite{8556482} & derived closed-form expressions of lower bound of SOP and their asymptotic behavior over the $\alpha-\eta-\mu$ \& $\alpha-\kappa-\mu$ fading channels.\\ \cline{2-3}
 & Zeng \textit{et al.}\cite{8369134}, \quad Zhao \textit{et al.} \cite{8777365}& analyzed the secrecy metrics  over the \textbf{FTR} wiretap fading channels.\\ \hline
\multirow{3}{*}{ 2020}&Kong \textit{et al.}\cite{9105083} & proposed a unified secrecy analysis framework with the help of \textbf{MoG} distribution. \\ \cline{2-3}
& Sánchez \textit{et al.} \cite{snchez2020physical} & derived the closed-form expressions of SOP and ASC metrics over \textbf{shadowed} $\kappa-\mu$ fading channels. \\ \cline{2-3}
& Sánchez \textit{et al.} \cite{snchez2020secrecy} & derived the exact and asymptotic SOP behavior over \textbf{NWDP} fading channels. \\
\hline
\bottomrule[1.4pt]
\end{tabular}
\end{table}

\subsection{Secrecy bounds}
Due to the difficulty in deriving exact closed-form SOP and ASC expressions, the lower bound of the SOP and ergodic secrecy capacity are often regarded as effective alternatives. 
\subsubsection{Lower bound of SOP}
The exact SOP can be accurately approximated by its lower bound when (i) the given transmission rate tends to zero, i.e., $R_t \rightarrow 0$; and (ii) Eve is closely located to Alice, which can be physically interpreted as Eve having an extremely high average received SNR, i.e., $\bar{\gamma}_E \rightarrow \infty$. In this context, the lower bound of SOP can be computed as 
\begin{equation}
\mathcal{P}_{out}^L  = \int_0^\infty F_B\left(2^{R_t}\gamma\right) f_E(\gamma) d\gamma. 
\end{equation}
Such an alternative has been widely investigated (see references \cite{7094262,7313027,7374839,8354927,LongFisherF}), and was shown to provide a fairly tight approximation. 
\subsubsection{Ergodic secrecy capacity}
To overcome the intractability of obtaining a closed-form ASC expression, the ergodic secrecy capacity is a widely adopted alternative to measure the average ability of secrecy transmission
over fading channels in open literature \cite{8850067}. For instance, in \cite{6805139}, the authors investigated the ergodic secrecy rate of downlink multiple-input multiple-output (MIMO) systems with limited CSI feedback. Moreover in \cite{LongIWCMC2018}, we have investigated the upper and lower bounds of the ergodic secrecy capacity of MIMO systems where zero-forcing (ZF) beamforming at Alice and ZF detectors at Bob and Eve are exploited.
\subsection{Secrecy Analysis Tools}
With the above in mind and under the assumption that the main and wiretap channels undergo independent fading conditions, this subsection will present three useful tools used to assist the secrecy analysis.   
\subsubsection{Mixture Gamma (MG) distribution}
According to \cite{6059452,7460203}, the instantaneous received SNR $\gamma$ over wireless Rayleigh, Nakagami-$m$, NL, $\kappa-\mu$, Hoyt, $\eta-\mu$, Rician, $\mathcal{K}$, $\mathcal{K}_G$, $\kappa-\mu$/gamma, $\eta-\mu$/gamma, and $\alpha-\mu$/gamma fading channels can be reformulated using the MG distribution, whereas the probability density function (PDF) and cumulative distribution function (CDF) of the instantaneous received SNR $\gamma$ are denoted as $f(\gamma)$ and $F(\gamma)$. 
\begin{subequations} 
\begin{equation}
f(\gamma) = \sum \limits_{l=1}^{L} \alpha_{l}\gamma^{\beta_{l} -1}\exp(-\zeta_{l}\gamma),
\end{equation}
\begin{equation}
F(\gamma) = \sum \limits_{l=1}^{L} \alpha_{l} \zeta_{l}^{-\beta_{l}} \Upsilon(\beta_{l},\zeta_{l}\gamma),
\end{equation}
\end{subequations}
where $L$ is the number of terms in the mixture, while $\alpha_{l}, \beta_{l}$, and $\zeta_{l}$ are the parameters of the $l$th gamma component. $\Upsilon(\cdot,\cdot)$ is the upper incomplete gamma function.

The work in \cite{7342908} used the MG distribution to assist the information-theoretic secrecy analysis. Motivated by \cite{8672771}, the secrecy metrics over the FTR and Málaga turbulence fading channels \cite{8369134,7859265} can be similarly derived using the MG distribution.
\subsubsection{Mixture of Gaussian (MoG) distribution} 
Based on the unsupervised expectation-maximization (EM) learning algorithm, the MoG distribution is essentially beneficial when the characteristics of the fading channel are unavailable. In \cite{7336572}, the authors modeled the RL, NL, $\eta-\mu$, $\kappa-\mu$, and shadowed $\kappa-\mu$ fading channels using the MoG distribution. The findings of \cite{7336572} showcases that the MoG distribution is especially advantageous to approximate any arbitrarily shaped non-Gaussian density, and can accurately model both composite and non-composite channels in a simple expression.

Assuming $\gamma$ follows the MoG distribution, its PDF and CDF are written as
\begin{subequations} 
\begin{equation}
f(\gamma) = \sum \limits_{l=1}^{C} \frac{w_l}{\sqrt{8\pi \bar{\gamma}}\eta_l \sqrt{\gamma} }\exp\left(-\frac{(\sqrt{\gamma/\bar{\gamma}} - \mu_l)^2}{2\eta_l^2}\right),
\end{equation}
\begin{equation}
F(\gamma) = \sum \limits_{l=1}^{C}w_l \Phi\left(\frac{\sqrt{ \gamma/\bar{\gamma}} - \mu_l}{\eta_l} \right),
\end{equation}
\end{subequations}
 where $C$ represents the number of Gaussian components. $w_l>0$, $\mu_l$, and $\eta_l$ are the $l$th mixture component's weight, mean, and variance with $\sum_l^C w_l = 1$, $\Phi(x)$ is the CDF of the standard normal distribution.
\subsubsection{Fox's $H$-function distribution}
 For known fading characteristics, the Fox's $H$-function distribution is a general and flexible tool that can model the instantaneous received SNR. It is reported in \cite{6226905,7089281,6550886,8706707} that many well-known distributions in the literature, e.g., Rayleigh, Exponential, Nakagami-$m$, Weibull, $\alpha-\mu$, gamma, Fisher-Snedecor $\mathcal{F}$, Chi-square, cascaded Rayleigh/Nakagami-$m$/$\alpha-\mu$, gamma-gamma, Málaga, $\mathcal{K}_G$, EGK, etc., can be represented using Fox's $H$-function distribution. For examples, interested readers are suggested to refer to Table. \ref{table_Fox}.

Assuming $\gamma$ follows Fox's $H$-function distribution, its PDF and CDF are given by 
\begin{subequations}
\begin{equation}
f(\gamma) = \mathcal{K} H_{p,q}^{m,n} \left[ {\mathcal{C} \gamma \left| {\begin{array}{*{20}c}
    {(a_1,A_1),\cdots,(a_p,A_p)}   \\
   {(b_1,B_1),\cdots,(b_q,B_q)}  \\
\end{array}} \right.} \right],
\end{equation}
\begin{equation}
\begin{split}
&F(\gamma) = \frac{\mathcal{K}}{\mathcal{C}} \\
&\times H_{p+1,q+1}^{m,n+1} \hspace{-1ex} \left[ {\mathcal{C} \gamma \left| \hspace{-1.2ex} {\begin{array}{*{20}c}
    {(1,1),(a_1+A_l,A_1),\cdots,(a_p+A_p,A_p)}   \\
   {(b_1+B_1,B_1),\cdots,(b_q+B_q,B_q),(0,1)}  \\
\end{array}} \right.}\hspace{-1.5ex} \right],
\end{split}
\end{equation}
\end{subequations}
where $H_{p,q}^{m,n}[.]$ is the univariate Fox's $H$-function \cite[Eq. (8.4.3.1)]{prudnikov1990integrals}, $\mathcal{K} > 0$ and $\mathcal{C}$ are constants such that $\int_0^\infty f (\gamma) d \gamma = 1$. $A_i > 0$ for $i=1,\cdots, p$, $B_l > 0$ for $l = 1, \cdots, q$, $ 0 \le m \le q$, and $0 \le n \le p$. For notational convenience, let $\mathfrak{a} = (a_1,\cdots,a_p)$, $\mathscr{A}= (A_1,\cdots,A_p)$, $\mathfrak{b} = (b_1,\cdots,b_q)$, and $\mathscr{B} = (B_1,\cdots,B_q)$. Thus, hereafter the Fox's $H$-function is denoted as $\mathcal{H}_{p,q}^{m,n}(\mathcal{K}, \mathcal{C}, \mathfrak{a}, \mathscr{A}, \mathfrak{b}, \mathscr{B})$.

\begin{table*}[htb!]
\renewcommand{\arraystretch}{1.5}
\caption{\textsc{Fox’s $H$-equivalents of typical and generalized statistical models}}
\label{table_Fox}
\centering
\begin{tabular}{|c|c|c|c|c|c|c|c|}
\arrayrulecolor{tabcolor}
\toprule[1.4pt] \hline
\rowcolor{gray}
\textbf{Fading model}& $m$ $n$ $p$ $q$&$\mathcal{K}$& $\mathcal{C}$& $\mathfrak{a}$&$\mathfrak{b}$&$\mathscr{A}$&$\mathscr{B}$ \\ \hline
Rayleigh & 1 0 0 1&$\frac{1}{\bar{\gamma}}$ &$\frac{1}{\bar{\gamma}}$ &- &0 & -& 1 \\ \hline
Nakagami-$m$ & 1 0 0 1&$\frac{m}{\Gamma(m)\bar{\gamma}}$ &$\frac{m}{\bar{\gamma}}$ &- &$m -1 $ &- &1 \\ \hline
Weibull & 1 0 0 1& $\frac{\Gamma(1+\frac{2}{\alpha})}{\bar{\gamma}}$ &$\frac{\Gamma(1+\frac{2}{\alpha})}{\bar{\gamma}}$ &- & $1 - \frac{2}{\alpha}$&- &$\frac{2}{\alpha}$ \\ \hline
$\alpha-\mu$ & 1 0 0 1& $\frac{\Gamma(\mu + \frac{2}{\alpha})}{\Gamma(\mu)^2\bar{\gamma}}$&$\frac{\Gamma(\mu+\frac{2}{\alpha})}{\Gamma(\mu)\bar{\gamma}}$ & -&$\mu - \frac{2}{\alpha}$ &- &$\frac{2}{\alpha}$ \\ \hline
Maxswell& 1 0 0 1& $\frac{3}{\sqrt{\pi}\bar{\gamma}}$ &$\frac{3}{2\bar{\gamma}}$ &- &$\frac{1}{2}$ &- &1 \\ \hline
$N*(\alpha-\mu)$ & $N$ 0 0 $N$ &$\prod \limits_{i=1}^N\frac{\Gamma(\mu_i + \frac{2}{\alpha_i})}{\Gamma(\mu_i)^2\bar{\gamma}} $& $\prod \limits_{i=1}^N\frac{\Gamma(\mu_i + \frac{2}{\alpha_i})}{\Gamma(\mu_i)\bar{\gamma}} $& -&$(\mu_1 - \frac{2}{\alpha_1},\cdots,\mu_N - \frac{2}{\alpha_N})$ &- &$(\frac{2}{\alpha_1},\cdots, \frac{2}{\alpha_N})$ \\ \hline
Fisher-Snedecor $\mathcal{F}$ & 1 1 1 1&$ \frac{m}{m_s\bar{\gamma} \Gamma(m)\Gamma(m_s)}$& $ \frac{m}{m_s\bar{\gamma} }$&$-m_{s}$ &1 &$m -1$ &1 \\ \hline
$\mathcal{K}_G$ & 2 0 0 2&$\frac{m_l m_{sl}}{\Gamma(m_l)\Gamma(m_{sl})\bar{\gamma}}$ & $ \frac{m_1m_2}{\bar{\gamma}}$& -&$(m_l-1 ,m_{s1} - 1)$ & -& $(1,1)$\\ \hline
EGK & 2 0 0 2&$\frac{\Gamma(m + \frac{1}{\xi}) \Gamma(m_s + \frac{1}{\xi_s})}{\bar{\gamma}\Gamma(m)^2\Gamma(m_s)^2}$ & $\frac{\Gamma(m + \frac{1}{\xi}) \Gamma(m_s + \frac{1}{\xi_s})}{\bar{\gamma}\Gamma(m)\Gamma(m_s)}$ &- &$(m - \frac{1}{\xi}, m_{s} - \frac{1}{\xi_{s}})$ & -& $(\frac{1}{\xi}, \frac{1}{\xi_{s}})$ \\ \hline
\bottomrule[1.4pt]
\end{tabular}
\end{table*}

\begin{figure*}[!t] 
\centering
\subfloat[$\mathcal{P}_{nz}^{MG}$]{\includegraphics[width=0.67\columnwidth]{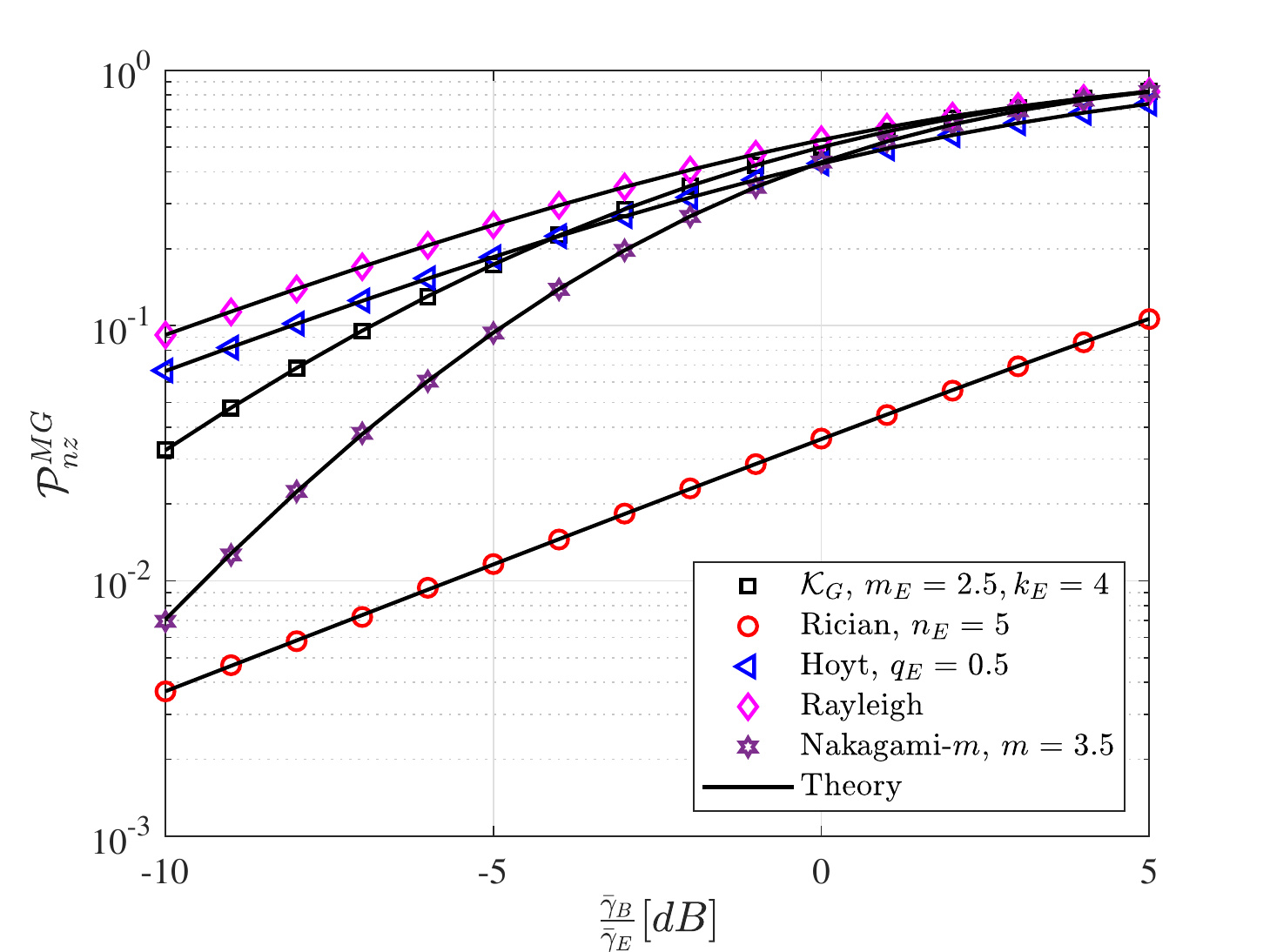}} 
\hfill
\subfloat[$\mathcal{P}_{nz}^{MoG}$]{\includegraphics[width=0.67\columnwidth]{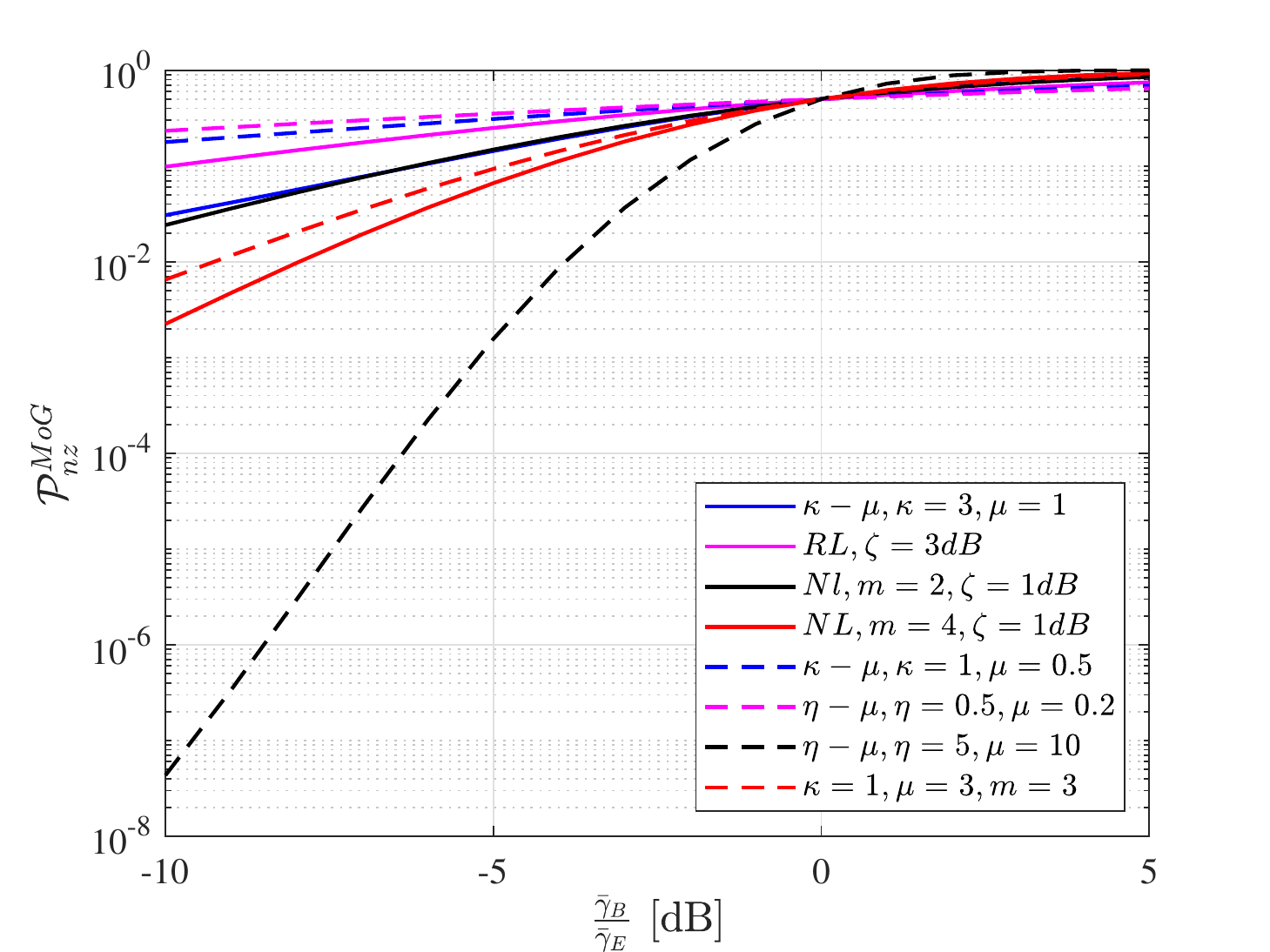}} 
\hfill
\subfloat[$\mathcal{P}_{nz}^{Fox}$]{\includegraphics[width=0.66\columnwidth]{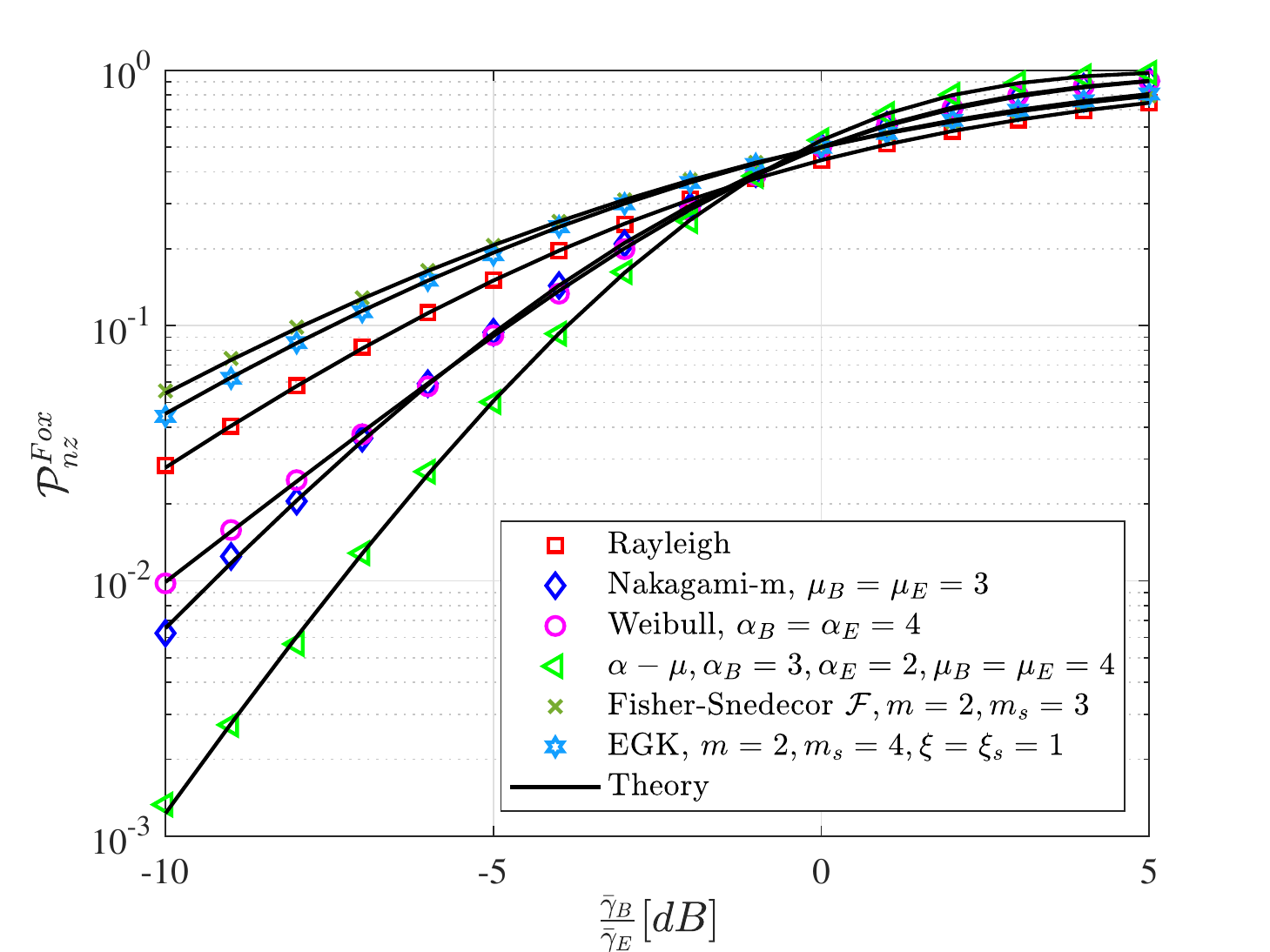}} 
\hfill
\caption{Illustration of $\mathcal{P}_{nz}$ versus $\frac{\bar{\gamma}_B}{\bar{\gamma}_E}$ using the MG, MoG, and Fox's $H$-function distributions when $\bar{\gamma}_E$ = 0dB, (a) the main channel undergoes $\mathcal{K}_G$ ($m_B = 2.5, k_B = 4$) fading while the wiretap channel experiences $\mathcal{K}_G$, Rician, Hoyt, Rayleigh, and Nakagami-$m$ ($m = 3.5$) fading; (b) the main and wiretap channels undergo same fading while using the MoG distribution; and (c) the main and wiretap channels undergo same fading while using the Fox's $H$-function distribution.}
\label{PNZ_MG_MoG_fig} 
\end{figure*}

For the purposes of comparing the secrecy analysis by using the three aforementioned approaches, the PNZ metric is taken as an example. Provided that the main and wiretap links undergo the same fading conditions, the PNZ expressions are given in terms of the Gauss Hypergeometric function \cite[Eq. (7)]{8672771}, error function \cite[Eq. (9)]{9105083}, and Fox's $H$-function \cite[Eq. (16)]{8706707}. In Fig. \ref{PNZ_MG_MoG_fig}, we plotted the PNZ performance versus $\bar{\gamma}_B$ for different fading channel models. Their tightness and accuracy have already been individually presented and confirmed in \cite{8672771,9105083,8706707}. 

\emph{Remark:} Conclusively speaking, the MG, MoG, and Fox's $H$-function distributions have demonstrated their feasibility and applicability when analyzing secrecy metrics. They all are valid when the main channel and wiretap channel are subjected to different wireless channels as shown in Fig. \ref{PNZ_MG_MoG_fig}. (a). 
\textit{\textbf{Note that the three aforesaid solutions are unfeasible when the main and wiretap channels are correlated}}.
\subsection{Outdated \& imperfect \& correlated CSI}
The aforementioned works mainly focus on the scenario that perfect CSI is available at all parties. Such an assumption is unrealistic, since in practice outdated CSI and imperfect CSI are the general cases due to the time varying nature of wireless channels and the channel estimation errors. 

In \cite{6496988}, the effects of outdated CSI on secrecy performance was investigated over multiple-input single-output (MISO) systems when the transmit antenna selection (TAS) scheme is applied at Alice. The obtained analytical results show that the diversity gain of using multiple antenna techniques cannot be achieved when the CSI is outdated during the TAS process. Later on in \cite{hu2015off}, Hu \textit{et al.} adopted the on-off-based transmission scheme at Alice to efficiently take advantage of the useful information in the outdated CSI. Alice does transmission only when she has a better link to Bob compared with that to Eve. Perfect knowledge of the main and wiretap channel CSI are always favorable, but the existence of noise in the channel estimation process makes it an unrealistic assumption. The impacts of imperfect CSI have been widely explored in diverse research topics, e.g., imperfect CSI in the artificial-noise-assisted training and communication \cite{7605496}, imperfect CSI with an active full-duplex eavesdropper \cite{7881216}, imperfect CSI in a mixed RF/FSO system \cite{8358703}, etc.

Apart from the above two scenarios, the correlation between the main channel and wiretap channel is also attracting a growing body of research interests. In \cite{5730598}, Jeon \textit{et al.} explored the secrecy capacity bounds considering correlated Rayleigh fading channels. The results quantitatively showcased how much of secrecy capacity is lost due to channel correlation. In continuation of this work, the secrecy
analysis exploration over correlated Nakagami-$m$, correlated $\alpha-\mu$, correlated shadowed $\kappa-\mu$ fading channels can be respectively found in \cite{8727439,8746270,8801940}.
\section{Secrecy Enhancement Approaches} \label{Sec4}
The essence of PLS is to utilize the impairments (e.g., fading, noise, interference, and path diversity) of wireless channels to enhance secrecy. In this section, we mainly focus on comparing the existing secrecy enhancement techniques suitable for classical wiretap channels. 
\subsection{On-off transmission scheme}
Considering imperfect channel estimation, He and Zhou in \cite{6623091} first proposed the on-off transmission scheme to improve the reliability and security performance. The principle of on-off transmission lies in the comparison between the estimated instantaneous SNRs at Bob ($\hat{\gamma}_B$) and Eve ($\hat{\gamma}_E$) and two given corresponding thresholds i.e., $\mu_B$ and $\mu_E$, to be specific, only when $\hat{\gamma}_B \ge \mu_B$ and $\hat{\gamma}_E \le \mu_E $, the "on" mode at Alice is activated, otherwise, Alice is in "off" mode. The on-off transmission scheme is alternatively an appealing candidate to allow the SOP metric be arbitrarily small. Building on He's work, the on-off transmission is widely investigated in the following secrecy works \cite{hu2015off,7926454,8110637,8541120}, where the imperfect CSI, outdated CSI, and correlated CSI are considered.
\subsection{Jamming approach} 
Assuming the transmitter has more antennas than the eavesdropper, Goel and Negi proposed the concept of artificial noise (AN) in \cite{4543070}. The principle behind AN is that the transmitter uses some of its available power to generate AN to confuse passive eavesdroppers. Consequently, Wang \textit{et al.} in \cite{6841050} proposed the artificial fast fading (AFF) secrecy enhancement scheme, where the randomized beamforming is employed at the transmitter in order to `upgrade' the main channel to an AWGN one and degrade the wiretap channel to a fast fading channel.

Different from the aforementioned transmitting beamforming-based techniques, namely AN and AFF, the quality of the wiretap link is further degraded by allocating part of the transmitting resources (i.e., power or antennas) at the transmitter, specifically to Eve. Based on the surveys in \cite{7324413,8088548}, one can conclude that jamming is another useful means to further enhance PLS. Considering the three-node wiretap fading channel, jamming can be realized by a full-duplex Bob, where Bob would receive signals from Alice and send jamming signals (i.e., noise) to Eve in order to reduce Eve's received SNR quality \cite{8355793}. Bob and Eve usually only act purely as a legitimate receiver or an illegitimate eavesdropper. However, practically speaking, they might behave with multiple roles. For instance, in \cite{7881216}, an active eavesdropper operates in full-duplex mode so that it can send jamming signals to degrade the legitimate receiver's SNR, while in \cite{6051523,6774842}, an untrustworthy relay works as a relay and eavesdropper simultaneously in a bidirectional cooperative network.

\subsection{Antenna selection technique}
In multiple-antenna systems, TAS is seen as an effective way for reducing hardware complexity whilst boosting diversity benefits. In \cite{6574305,wang2014secure,6328208,zhu2015secrecy,7118654,8541120,8801940,8734105}, TAS is deployed as a secrecy enhancement solution in MIMO systems. There exist three kinds of TAS schemes: (i) the antenna that maximizes the output instantaneous SNR at Bob is selected (see \cite{6574305,wang2014secure}); (ii) more than one single antenna are selected (see \cite{6328208}); and (iii) a general order of antenna are selected (see \cite{7118654}).

Different from the works \cite{6574305,wang2014secure,6328208,zhu2015secrecy} assuming that the multi-antenna channels are independent, quite recently, Si \textit{et al.} considered antenna correlation in \cite{8541120}, where the exact and asymptotic SOP are derived with consideration of three diversity combining schemes, namely maximal ratio combining (MRC), selection combining (SC), and equal gain combining (EGC) at Bob. This work was extended in \cite{8801940}, where continuously took consideration of the joint antenna and channel correlation, while the relationship between the correlation and the SOP is analytically established.
\subsection{Protected zones}
Protected zones (equivalently, secrecy region) means a geometrical region (see \cite{6840982,7247765}), defined as the legitimate receiver's locations having a certain guaranteed level of secrecy, or an area where the set of ordered nodes can safely communicate with typical destination, for a given secrecy
outage constraint \cite{7942069,8477185}. 
\section{Concluding remarks} \label{Sec5}
In this paper, we have comprehensively summarized the development of PLS over various wiretap fading channels. Based on the channel characteristics, the existing secrecy analysis works are classified into four categories: (i) small-scale fading; (ii) large-scale fading; (iii) cascaded fading; and (iv) composite fading. By presenting and comparing some major information-theoretic secrecy analysis works, we found three useful and effective approaches, i.e., the MG, MoG, and Fox's $H$-function distributions, which are introduced to simplify the secrecy analysis. The three aforesaid approaches are beneficial since they can largely cover the existing fading models when the main and wiretap channels experience independent fading condition. In addition, we presented the existing secrecy enhancement techniques deployed for Wyner's wiretap channel model, including on-off transmission, jamming approach, TAS technique, and protected zones.

\bibliographystyle{IEEEtran}
\bibliography{FoxHfadingChannel}
\end{document}